\documentclass[twocolumn,showpacs,preprintnumbers,amsmath,amssymb,prl,superscriptaddress]{revtex4}

\usepackage{graphicx} % Include figure files
\usepackage{dcolumn}  % Align table columns on decimal point
\usepackage{bm}       % bold math 
\usepackage{alltt}
\usepackage{calc,rotating,xspace}
\usepackage{pifont}

\newcommand{\BsBsbar}{\ensuremath{B^0_s}-\ensuremath{\overline{B}^0_s}}
\newcommand{\Bs}     {\ensuremath{B_s^0}}
\newcommand{\Bsbar}  {\ensuremath{\overline{B}^0_s}} %IJK
\newcommand{\bs}     {\ensuremath{B_s}} %added by IJK

\newcommand{\Dsp}    {\ensuremath{D_s^+}}
\newcommand{\Bd}     {\ensuremath{B^0}}

\newcommand{\keV}    {\ensuremath{\rm ke\kern -0.1em V}}
\newcommand{\MeV}    {\ensuremath{\rm Me\kern -0.1em V}}

\newcommand{\MeVcc}  {\ensuremath{\MeV\!/\mathit{c}^2}}
\newcommand{\GeV}    {\ensuremath{\rm Ge\kern -0.1em V}}
\newcommand{\TeV}    {\ensuremath{\rm Te\kern -0.1em V}}

\newcommand{\GeVcc}  {\ensuremath{\GeV\!/\mathit{c}^2}}%%%added by IJK

\newcommand{\ifb}    {\ensuremath{\mathrm{fb^{-1}}}}
\newcommand{\ips}    {\ensuremath{\mathrm{ps^{-1}}}}

\newcommand{\dms}    {\ensuremath{\Delta m_s}}
\newcommand{\dmd}    {\ensuremath{\Delta m_d}}
\newcommand{\Vtd}    {\ensuremath{V_{td}}}
\newcommand{\Vts}    {\ensuremath{V_{ts}}}
\newcommand{\VtsVtd} {\ensuremath{|\Vts/\Vtd|}}
\newcommand{\VtdVts} {\ensuremath{|\Vtd/\Vts|}}
\newcommand{\Lrat}   {\ensuremath{\Lambda}}

\newcommand{\sye}[1]{\ensuremath{~\pm #1}}
\newcommand{\ase}[2]{\ensuremath{^{~+ #1}_{~- #2}}}
\newcommand{\massRatio} {\ensuremath{m_{\Bd}/m_{\Bs} = 0.98390}}
\newcommand{\deltaMdPdg}{\ensuremath{\dmd = 0.505\sye{0.005}\,\ips}}
\newcommand{\xiLat}     {\ensuremath{\xi = 1.21\ase{0.047}{0.035}}}
\newcommand{\deltaMsResult}%
{\ensuremath{\dms =
  17.31\ase{0.33}{0.18}~({\rm stat.})\sye{0.07}~({\rm syst.})\,\ips}}
\newcommand{\VtsResult}%
{\ensuremath{\VtsVtd =
  4.791\ase{0.152}{0.190}~({\rm stat. + syst.})}}
\newcommand{\VtdResult}%
{\ensuremath{\VtdVts =
  0.208\ase{0.001}{0.002}{\rm (exp.)}\ase{0.008}{0.006}{\rm(theo.)}}}

\begin{document}
%%%%%%%%%%%%%%%%%%%%%%%%%%%%%%%%%%%%%%%%%%%%%%%%%%%%%%%%%%%%%%%%%%%%%%%%%%%%%%%%

%% comment OUT next line for single-column-with-line-numbers
%%%{\hfill \today\ --- Version $3.0$}

%%%%%%%%%%%%%%%%%%%%%%%%%%%%%%%%%%%%%%%%%%%%%%%%%%%%%%%%%%%%%%%%%%%%%%%%%%%%%%%%
% abstract and title
%%%%%%%%%%%%%%%%%%%%%%%%%%%%%%%%%%%%%%%%%%%%%%%%%%%%%%%%%%%%%%%%%%%%%%%%%%%%%%%%

%\begin{LARGE}
%\center{\bf \boldmath Measurement of the {\BsBsbar} Oscillation Frequency}
%\end{LARGE}

\title{\boldmath Measurement of the {\BsBsbar} Oscillation Frequency}

%% UNcomment next line for single-column-with-line-numbers
%\maketitle

%%%\input{January_2006_Authors.tex}
\affiliation{Institute of Physics, Academia Sinica, Taipei, Taiwan 11529, Republic of China} 
\affiliation{Argonne National Laboratory, Argonne, Illinois 60439} 
\affiliation{Institut de Fisica d'Altes Energies, Universitat Autonoma de Barcelona, E-08193, Bellaterra (Barcelona), Spain} 
\affiliation{Baylor University, Waco, Texas  76798} 
\affiliation{Istituto Nazionale di Fisica Nucleare, University of Bologna, I-40127 Bologna, Italy} 
\affiliation{Brandeis University, Waltham, Massachusetts 02254} 
\affiliation{University of California, Davis, Davis, California  95616} 
\affiliation{University of California, Los Angeles, Los Angeles, California  90024} 
\affiliation{University of California, San Diego, La Jolla, California  92093} 
\affiliation{University of California, Santa Barbara, Santa Barbara, California 93106} 
\affiliation{Instituto de Fisica de Cantabria, CSIC-University of Cantabria, 39005 Santander, Spain} 
\affiliation{Carnegie Mellon University, Pittsburgh, PA  15213} 
\affiliation{Enrico Fermi Institute, University of Chicago, Chicago, Illinois 60637} 
\affiliation{Joint Institute for Nuclear Research, RU-141980 Dubna, Russia} 
\affiliation{Duke University, Durham, North Carolina  27708} 
\affiliation{Fermi National Accelerator Laboratory, Batavia, Illinois 60510} 
\affiliation{University of Florida, Gainesville, Florida  32611} 
\affiliation{Laboratori Nazionali di Frascati, Istituto Nazionale di Fisica Nucleare, I-00044 Frascati, Italy} 
\affiliation{University of Geneva, CH-1211 Geneva 4, Switzerland} 
\affiliation{Glasgow University, Glasgow G12 8QQ, United Kingdom} 
\affiliation{Harvard University, Cambridge, Massachusetts 02138} 
\affiliation{Division of High Energy Physics, Department of Physics, University of Helsinki and Helsinki Institute of Physics, FIN-00014, Helsinki, Finland} 
\affiliation{University of Illinois, Urbana, Illinois 61801} 
\affiliation{The Johns Hopkins University, Baltimore, Maryland 21218} 
\affiliation{Institut f\"{u}r Experimentelle Kernphysik, Universit\"{a}t Karlsruhe, 76128 Karlsruhe, Germany} 
\affiliation{High Energy Accelerator Research Organization (KEK), Tsukuba, Ibaraki 305, Japan} 
\affiliation{Center for High Energy Physics: Kyungpook National University, Taegu 702-701, Korea; Seoul National University, Seoul 151-742, Korea; and SungKyunKwan University, Suwon 440-746, Korea} 
\affiliation{Ernest Orlando Lawrence Berkeley National Laboratory, Berkeley, California 94720} 
\affiliation{University of Liverpool, Liverpool L69 7ZE, United Kingdom} 
\affiliation{University College London, London WC1E 6BT, United Kingdom} 
\affiliation{Centro de Investigaciones Energeticas Medioambientales y Tecnologicas, E-28040 Madrid, Spain} 
\affiliation{Massachusetts Institute of Technology, Cambridge, Massachusetts  02139} 
\affiliation{Institute of Particle Physics: McGill University, Montr\'{e}al, Canada H3A~2T8; and University of Toronto, Toronto, Canada M5S~1A7} 
\affiliation{University of Michigan, Ann Arbor, Michigan 48109} 
\affiliation{Michigan State University, East Lansing, Michigan  48824} 
\affiliation{Institution for Theoretical and Experimental Physics, ITEP, Moscow 117259, Russia} 
\affiliation{University of New Mexico, Albuquerque, New Mexico 87131} 
\affiliation{Northwestern University, Evanston, Illinois  60208} 
\affiliation{The Ohio State University, Columbus, Ohio  43210} 
\affiliation{Okayama University, Okayama 700-8530, Japan} 
\affiliation{Osaka City University, Osaka 588, Japan} 
\affiliation{University of Oxford, Oxford OX1 3RH, United Kingdom} 
\affiliation{University of Padova, Istituto Nazionale di Fisica Nucleare, Sezione di Padova-Trento, I-35131 Padova, Italy} 
\affiliation{LPNHE, Universite Pierre et Marie Curie/IN2P3-CNRS, UMR7585, Paris, F-75252 France} 
\affiliation{University of Pennsylvania, Philadelphia, Pennsylvania 19104} 
\affiliation{Istituto Nazionale di Fisica Nucleare Pisa, Universities of Pisa, Siena and Scuola Normale Superiore, I-56127 Pisa, Italy} 
\affiliation{University of Pittsburgh, Pittsburgh, Pennsylvania 15260} 
\affiliation{Purdue University, West Lafayette, Indiana 47907} 
\affiliation{University of Rochester, Rochester, New York 14627} 
\affiliation{The Rockefeller University, New York, New York 10021} 
\affiliation{Istituto Nazionale di Fisica Nucleare, Sezione di Roma 1, University of Rome ``La Sapienza," I-00185 Roma, Italy} 
\affiliation{Rutgers University, Piscataway, New Jersey 08855} 
\affiliation{Texas A\&M University, College Station, Texas 77843} 
\affiliation{Istituto Nazionale di Fisica Nucleare, University of Trieste/\ Udine, Italy} 
\affiliation{University of Tsukuba, Tsukuba, Ibaraki 305, Japan} 
\affiliation{Tufts University, Medford, Massachusetts 02155} 
\affiliation{Waseda University, Tokyo 169, Japan} 
\affiliation{Wayne State University, Detroit, Michigan  48201} 
\affiliation{University of Wisconsin, Madison, Wisconsin 53706} 
\affiliation{Yale University, New Haven, Connecticut 06520} 
\author{A.~Abulencia}
\affiliation{University of Illinois, Urbana, Illinois 61801}
\author{D.~Acosta}
\affiliation{University of Florida, Gainesville, Florida  32611}
\author{J.~Adelman}
\affiliation{Enrico Fermi Institute, University of Chicago, Chicago, Illinois 60637}
\author{T.~Affolder}
\affiliation{University of California, Santa Barbara, Santa Barbara, California 93106}
\author{T.~Akimoto}
\affiliation{University of Tsukuba, Tsukuba, Ibaraki 305, Japan}
\author{M.G.~Albrow}
\affiliation{Fermi National Accelerator Laboratory, Batavia, Illinois 60510}
\author{D.~Ambrose}
\affiliation{Fermi National Accelerator Laboratory, Batavia, Illinois 60510}
\author{S.~Amerio}
\affiliation{University of Padova, Istituto Nazionale di Fisica Nucleare, Sezione di Padova-Trento, I-35131 Padova, Italy}
\author{D.~Amidei}
\affiliation{University of Michigan, Ann Arbor, Michigan 48109}
\author{A.~Anastassov}
\affiliation{Rutgers University, Piscataway, New Jersey 08855}
% ...
\collaboration{CDF Collaboration}
\noaffiliation

\begin{abstract}
 We present the first measurement of the
 {\BsBsbar} oscillation frequency {\dms}.
 We use 1~{\ifb} of data from $p\bar{p}$ collisions at $\sqrt{s}=1.96~\TeV$
 collected with the CDF\,II detector at the Fermilab Tevatron.
 The sample contains signals of 3,600 fully reconstructed hadronic {\bs}
 decays and 37,000 partially reconstructed semileptonic {\bs} decays.
% We measure the probability as a function of proper decay time that the
% {\bs} decays with the same, or opposite, flavor as the flavor at
% production, which is determined using opposite-side and same-side flavor
% identification methods.
% We find a signal consistent with {\BsBsbar} oscillations.
 We measure the probability as a function of proper decay time that the
 {\bs} decays with the same, or opposite, flavor as the flavor at
 production, and we find a signal consistent with {\BsBsbar} oscillations.
 The probability that random fluctuations could produce a comparable signal 
 is 0.2\%.
 Under the hypothesis that the signal is due to {\BsBsbar} oscillations, we
 measure {\deltaMsResult} and determine {\VtdResult}.
\end{abstract}

%% add PACS numbers


%% comment OUT next line for single-column-with-line-numbers
\maketitle

%% UNcomment next line for single-column-with-line-numbers
%\clearpage

%%%%%%%%%%%%%%%%%%%%%%%%%%%%%%%%%%%%%%%%%%%%%%%%%%%%%%%%%%%%%%%%%%%%%%%%%%%%%%%%
% main body of text
%%%%%%%%%%%%%%%%%%%%%%%%%%%%%%%%%%%%%%%%%%%%%%%%%%%%%%%%%%%%%%%%%%%%%%%%%%%%%%%%

%%%\input{mainbody.tex}

%%%%%%%%%%%%%%%%%%%%%%%%%%%%%%%%%%%%%%%%%%%%%%%%%%%%%%%%%%%%%%%%%%%%%%%%%%%%%%%%
% introduction
%%%%%%%%%%%%%%%%%%%%%%%%%%%%%%%%%%%%%%%%%%%%%%%%%%%%%%%%%%%%%%%%%%%%%%%%%%%%%%%%
	
 Neutral $B$ mesons ($b\bar{q}$, with $q=d,s$ for $\overline{B}^0_d$,
 $\overline{B}^0_s$) oscillate from particle to antiparticle due to 
 flavor-changing weak interactions.
 The probability density $P_+$ ($P_-$) for a $\overline{B}^0_q$~meson
 produced at proper time $t=0$ to decay as a $\overline{B}^0_q$ ($B^0_q$)
 at time $t$ is given by
\begin{displaymath}
  P_{\pm}(t) = 
  \frac{\Gamma_q}{2} e^{-\Gamma_q t} \left [ 1\pm\cos(\Delta m_q t) \right]\,,
\end{displaymath}
 where $\Delta m_q$ is the mass difference between the two mass eigenstates
 $B^0_{q,H}$ and $B^0_{q,L}$~\cite{MIXING}, and $\Gamma_q$ is the decay width,
 which is assumed to be equal for the two mass eigenstates.
 The mass differences $\Delta m_d$ and $\Delta m_s$ can be used to determine
 the fundamental parameters $|{\Vtd}|$ and $|{\Vts}|$, respectively, 
% the two most poorly constrained elements of the Cabibbo-Kobayashi-Maskawa
% (CKM) matrix~\cite{CKM}.
 of the Cabibbo-Kobayashi-Maskawa (CKM) matrix~\cite{CKM}, which relates the
 quark mass eigenstates to the flavor eigenstates.
 This determination, however, has large theoretical uncertainties.
 A measurement of $\dms$ combined with $\deltaMdPdg$~\cite{PDG2006,DMD}
% The measurement $\deltaMdPdg$~\cite{PDG2006,DMD} combined with a
% measurement of $\dms$
 would determine the ratio {\VtdVts} with a significantly smaller theoretical
 uncertainty, contributing to a stringent test of the unitarity of the
 CKM matrix.
 Earlier attempts to measure $\dms$ have yielded a lower limit: 
 $\dms > 14.5~\ips$~\cite{PDG2006,DMS} at the 95\% confidence level~(C.L.).
 Recently the D\O\ Collaboration reported $17~\ips<\Delta m_s<21~\ips$
 at 90\% C.L.~\cite{D0-BSMIX-2006} using a large sample of semileptonic
 {\bs}~\cite{NOTATION} decays.
 
 In this letter we report a measurement of $\Delta m_s$ using data from
 $1~\ifb$ of $p\bar{p}$ collisions at $\sqrt{s}=1.96~\TeV$ collected by
 the CDF\,II detector at the Fermilab Tevatron.
% and we use our measurement to determine the ratio {\VtdVts}.
 We begin by reconstructing {\bs} decays in hadronic
 ($\Bsbar\to D^+_s\pi^-$, $D^+_s\pi^-\pi^+\pi^-$) and semileptonic
 ($\Bsbar\to D^{+(*)}_s\ell^-\bar{\nu}_\ell$, $\ell=e$ or $\mu$)
 decay modes using charged particles only~\cite{CHARGECONJUGATE}.
 Using the method of maximum likelihood, we extract the value of $\dms$
 from the probability density functions (PDFs) that describe
 the measured time development of {\bs}~mesons that decay with the same
 or opposite flavor as their flavor at production.
 The proper decay time for each {\bs} is calculated from the
 measured distance between the production and decay points,
 the measured momentum, and the {\bs} mass
 $m_{\bs}=5.3696\,\GeVcc$~\cite{PDG2006}.
 The {\bs} flavor ($b$ or $\bar{b}$) at decay is determined unambiguously
 by the charges of the decay products.

 To identify the flavor of the {\bs} at production,
 we use characteristics of $b$ quark production and
 fragmentation in $p\bar{p}$ collisions.
 At the Tevatron, the dominant $b$~quark production mechanisms produce
 $b\bar{b}$~pairs.
 The $b$ and $\bar{b}$ are expected to fragment independently into hadrons.
 In a simple model of fragmentation, a $b$~quark becomes a $\Bsbar$~meson
 when some of the energy of the $b$~quark is used to produce an $s\bar{s}$
 quark pair.
 The $b$ and the $\bar{s}$ bind to form a $\Bsbar$.
 The remaining $s$~quark may form a $K^-$.
 Similarly, a $\bar{b}$ that becomes a $\Bs$ is accompanied by a $K^+$.
 One of the two techniques used to identify the production flavor of the {\bs}
 is based on the charge of these kaons (same-side tag).
 The second technique uses the charge of the lepton from semileptonic decays
 or a momentum-weighted charge of the decay products of the second $b$~hadron
 produced in the collision (opposite-side tag).
%We identify the production flavor of the $\bs$ using two techniques: 
%the first is based on identifying the flavor of the second $b$~hadron
%produced in the collision (opposite-side tag).
%We use the charge of the lepton from semileptonic decays,
%or a momentum weighted charge of the decay products, of the second $b$~hadron.
%The second technique is based on the charge of identified kaons
%produced in association with the fragmentation of a $b$ quark to a
%$\overline{B}^0_s$ (same-side tag).
 
 The hadronic and semileptonic decay modes are complementary.
 Due to the large branching ratio, the semileptonic decays provide
 a tenfold advantage in signal rate at the cost of significantly worsened
 decay-time resolution due to the unmeasured $\nu$ momentum.
 Semileptonic decays dominate the sensitivity to oscillations at lower
 values of {\dms}.
 The fully reconstructed hadronic {\bs} decays have superior decay time
 resolution, and our large sample of these decays
 is the unique feature that makes CDF sensitive to much
 larger values of {\dms} than other experiments.

%%%%%%%%%%%%%%%%%%%%%%%%%%%%%%%%%%%%%%%%%%%%%%%%%%%%%%%%%%%%%%%%%%%%%%%%%%%%%%%%
% detector
%%%%%%%%%%%%%%%%%%%%%%%%%%%%%%%%%%%%%%%%%%%%%%%%%%%%%%%%%%%%%%%%%%%%%%%%%%%%%%%%

 The CDF\,II detector~\cite{DETECTOR_REFERENCE}
 consists of a magnetic spectrometer surrounded by electromagnetic and
 hadronic calorimeters and muon detectors~\cite{MUONS}.
 %Electron detection is described in~\cite{BSUBC_PSI_E}.
 The key features for this measurement include precision vertex determination
 provided by the seven-layer double-sided inner silicon strip
 detector~\cite{SVX, ISL} supplemented with a single-sided layer of
 silicon~\cite{L00}
 mounted directly on the beampipe at an average radius of 1.5\,cm.
 The 96-layer outer drift chamber~\cite{COT} is
 used for both precision tracking and $dE/dx$ particle identification.
 Time-of-flight (TOF) counters~\cite{TOF} located just outside the drift
 chamber are used to identify low momentum charged kaons.
% We use a cylindrical coordinate system ($r$, $\varphi$, $z$) with
% $z$ pointing along the momentum of the proton beam.
% Pseudorapidity is $\eta=-\ln\tan(\theta/2)$, where $\theta$ is the polar
% angle of the momentum vector with respect to the $z$~axis;
% the transverse momentum is $p_T = p\sin\theta$.

%%%%%%%%%%%%%%%%%%%%%%%%%%%%%%%%%%%%%%%%%%%%%%%%%%%%%%%%%%%%%%%%%%%%%%%%%%%%%%%%
% trigger and data sample
%%%%%%%%%%%%%%%%%%%%%%%%%%%%%%%%%%%%%%%%%%%%%%%%%%%%%%%%%%%%%%%%%%%%%%%%%%%%%%%%

 Charm and bottom hadrons are selected using a three-level trigger system
 that exploits the kinematics of production and decay, and the
 long lifetimes of $D$ and $B$~mesons.
 A crucial component of the trigger system for this measurement is the
 Silicon Vertex Trigger~\cite{SVT}, which selects
 events that contain $\Bsbar\to D^+_s\pi^-$ and $D^+_s\pi^-\pi^+\pi^-$ decays.
 The trigger configuration used to collect the heavy flavor data sample
 is described in~\cite{CDF-BR-BSDSPI}.

 To reconstruct {\Bsbar} candidates, we first select $D^+_s$ candidates.
 We use $D^+_s\rightarrow \phi\pi^+, K^{*}(892)^{0} K^+$, and
 $\pi^+\pi^-\pi^+$, with $\phi\rightarrow K^+K^-$
 and $K^{*0}\rightarrow K^+\pi^-$;
 we require that $\phi$ and $K^{*0}$ candidates be consistent
 with the known masses and widths~\cite{PDG2006} of these two resonances.
 %We do not use $dE/dx$ or TOF measurements for particle identification
 %to select these combinations.
 These $D^+_s$ candidates are combined with one or three additional
 charged particles
 to form $D^+_s\ell^-$, $D^+_s\pi^-$ or $D^+_s\pi^-\pi^+\pi^-$ candidates.
%remove new paragraph here
 The $D^+_s$ and other decay products of a {\Bsbar} candidate
 are constrained to originate from a common vertex in three dimensions.
 For the $K^{*}(892)^{0} K^+$ final state, we remove candidates that are
 consistent with the decay $D^+\rightarrow K^-\pi^+\pi^+$.
 We use a likelihood technique to identify
 muons~\cite{GIURGIU_THESIS} and electrons~\cite{BSUBC_PSI_E}.
% Leptons are identified using a likelihood technique.
% derived from the probability density functions (PDF) of various detector
% quantities. 
% These PDFs are determined using independent data samples.

 Backgrounds are suppressed by imposing
 a requirement on the minimum transverse momentum $p_T$~\cite{DEFINE_PT}
 of the {\Bsbar} and
 by requiring that the {\Bsbar} and $D^+_s$ decay vertices
 are displaced significantly from the $p\bar{p}$ collision position.
 We find signals of 3,600 hadronic {\bs} decays and
 37,000 semileptonic {\bs} decays.

%%%%%%%%%%%%%%%%%%%%%%%%%%%%%%%%%%%%%%%%%%%%%%%%%%%%%%%%%%%%%%%%%%%%%%%%%%%%%%%%
% mass and lifetime
%%%%%%%%%%%%%%%%%%%%%%%%%%%%%%%%%%%%%%%%%%%%%%%%%%%%%%%%%%%%%%%%%%%%%%%%%%%%%%%%

% Fig.~\ref{fig:m-excl-bs} is an example of a mass
% spectrum for hadronic decays.
 For the hadronic decays, the invariant mass distribution has a signal 
 centered close to
 $m_{\bs}=5.3696\,\GeVcc$ with a width of 14 to 20\,\MeVcc,
 depending on the decay mode. 
 Candidates with masses greater than 5.5\,$\GeVcc$ are used to construct
 PDFs for combinatorial background.
% We model small reflections from other $b$~hadron decays
% with Monte Carlo simulation.
 To remove contributions from $\Bsbar \rightarrow D_s^{*+} \pi^-$, $\Bsbar
 \rightarrow \Dsp \rho^-$ and semileptonic and other partially reconstructed
 decays, we require the mass of the decay candidates to be greater than
 5.3 $\GeVcc$.
 For semileptonic decays we take into account several
 background contributions, including $B$ meson decays to two charm mesons
 and real $D_s$~mesons associated with a false lepton.

 The decay time in the $\bs$ rest frame is
 $t = \kappa [L_{T} m_{\bs}/p_T]$, 
 where $L_{T}$ is the displacement of the $\bs$
 decay vertex with respect to the primary vertex
 projected onto the $\bs$ transverse momentum vector.
% in the case of hadronic
% decays, or onto the transverse momentum vector of the combination of the
% $D_s$~meson and the lepton in the case of semileptonic decays.
 The factor $\kappa$ corrects for missing momentum in the semileptonic decays
 ($\kappa=1$ for hadronic decays).
 To improve the decay-time resolution, we use event-by-event
 primary vertex position measurements when computing the $\bs$
 vertex displacement.
 The signal decay-time distribution is modeled with
% $\partial P(t_i) / \partial t
% = \Gamma_s \cdot e^{-\Gamma_s t'} \otimes {\cal G}(t' - t_i,
% \sigma_{t_i}) \cdot \varepsilon(t_i)$,
 $P(t_i,\sigma_{t_i})
 =  \varepsilon(t_i) \int \Gamma_s e^{-\Gamma_s t'} 
    {\cal G}(t' - t_i, \sigma_{t_i}) dt'$,
 where $t_i$ is the measured decay
 time of the $i$th candidate, $\Gamma_s$ is the $\bs$ decay width,
 ${\cal G}(x-\mu,\sigma)$ is a Gaussian distribution of the random variable
 $x$ with mean $\mu$ and width $\sigma$, and $\sigma_{t_i}$
 is the estimated candidate decay-time resolution.
% The function $\varepsilon(t)$ is the decay-time efficiency function that
 The decay-time efficiency function $\varepsilon(t)$
 describes trigger and selection biases on the decay-time distribution
 and is determined from Monte Carlo simulation.
% The effect of the trigger and selection criteria on the decay time
% distribution is described by an efficiency function $\varepsilon(t)$
% determined by Monte Carlo simulation.
 For semileptonic decays, the $\kappa$ distribution is determined from
 Monte Carlo simulation
 and is convoluted with the signal decay-time distribution.
 The missing transverse momentum from unreconstructed particles in the
 semileptonic decays is an important contribution to the decay-time resolution.
 To reduce this contribution and make optimal use of the semileptonic decays,
 we determine the $\kappa$ distribution
 as a function of the invariant mass of the $D_s\ell$ pair, $m_{D_s\ell}$.
 The r.m.s.~width of the $\kappa$ distribution is 3\% (20\%) for
 $m_{D_s\ell} = 5.2~\GeVcc$ ($3.0~\GeVcc$).
 %Understanding the decay time resolution is crucial.
 %It is measured using a large sample of prompt
 %$D$ mesons~\cite{DPRODUCTION_PRL}, combined with one or three tracks from
 %the primary vertex to mimic the signal topologies.

 We estimate the decay-time resolution $\sigma_{t_i}$ for each candidate
 using the measured track parameters and their estimated uncertainties.
 We calibrate this estimate using a large sample of prompt
 $D^+$~mesons~\cite{DPRODUCTION_PRL}, which we combine with one or three
 charged particles from the primary vertex to mimic signal topologies.
 For hadronic decays the average decay-time resolution is 87\,fs, which
 corresponds to one fifth of an oscillation period at the lower limit
 on $\dms$ (14.5\,{\ips}).
 For semileptonic decays, the decay-time resolution is worse due to decay
 topology and the missing momentum of unreconstructed decay products.
 For example, at $t=0$, $\sigma_t=100$\,fs (200\,fs) for 
 $m_{D_s\ell} = 5.2~\GeVcc$ ($3.0~\GeVcc$) and increases to
 $\sigma_t=115$\,fs (380\,fs) at $t=1.5$\,ps. 

%%%%%%%%%%%%%%%%%%%%%%%%%%%%%%%%%%%%%%%%%%%%%%%%%%%%%%%%%%%%%%%%%%%%%%%%%%%%%%%%
% flavor
%%%%%%%%%%%%%%%%%%%%%%%%%%%%%%%%%%%%%%%%%%%%%%%%%%%%%%%%%%%%%%%%%%%%%%%%%%%%%%%%

 The flavor of the {\bs} at production is determined using both
 opposite-side and same-side flavor tagging techniques.
 The effectiveness $Q\equiv\epsilon {\cal D}^2$ of these techniques
 is quantified with an efficiency $\epsilon$, the fraction of signal candidates
 with a flavor tag, and a dilution ${\cal D}\equiv 1-2w$,
 where $w$ is the probability that the tag is incorrect.
 
 Opposite-side tags infer the production flavor of the {\bs} from the
 decay products
 of the $b$~hadron produced from the other $b$~quark in the event. 
% The opposite-side flavor tags are based on identifying the flavor of
% the $b$~hadron produced from the other $b$~quark in the event to infer
% the {\bs} flavor at production.
 We use lepton ($e$ and $\mu$) charge
 and jet charge as tags, building on
 techniques developed for a CDF Run\,I measurement of
 {\dmd}~\cite{OWEN_PRD}.
 If both lepton and jet-charge tags are present, we use the lepton tag,
 which has a higher average dilution.
 
 The dilution of opposite-side flavor tags is expected to be independent of
 the type of $B$~meson that produces the hadronic or semileptonic decay.
 The dilution is measured in data using
 large samples of $B^-$, which do not change flavor, and $\overline{B}^0$,
 which can be used after accounting for their well-known oscillation frequency.
 The combined opposite-side tag effectiveness is $Q = 1.5\pm0.1\,\%$,
 where the uncertainty is dominated by the statistics of the control samples.
 
 Same-side flavor tags~\cite{SST} are based on the charges
 of associated particles produced in the fragmentation of the
 $b$~quark that produces the reconstructed {\bs}.
 In the simplest picture of fragmentation,
 a $\pi^+$ ($\pi^-$) accompanies the formation of a $B^-$ ($B^+$),
 a $\pi^-$ ($\pi^+$) acccompanies a $\overline{B}^0$ ($B^0$),
 and a $K^-$ ($K^+$) accompanies a {\Bsbar} ({\Bs}).
 In Run\,I, CDF established this method of production flavor identification
 in measurements of {\dmd}~\cite{PETAR} and the $\mathit{CP}$
 symmetry violating parameter $\sin (2\beta)$~\cite{CDF_SIN2BETA}.
 In this analysis,
 we use $dE/dx$~\cite{BSUBC_PSI_E} and time-of-flight 
 information in a combined particle identification likelihood to identify
 the kaons associated with {\bs} production.
% Tracks within a cone of radius $\Delta R =
% \sqrt{\Delta\varphi^2+\Delta\eta^2} = 0.7$ and $p_T>0.45$\,\GeVc\ are
% considered as same-side tag candidates, and the track with the
% largest kaon likelihood is selected as the tagging track.
 Tracks close in phase space to the $\bs$ candidate are
 considered as same-side kaon tag candidates, and the track with the
 largest kaon likelihood is selected as the tagging track.
 
 The performance of the same-side kaon tag for {\Bsbar} is expected to be
 different than for $B^-$ and $\overline{B}^0$.
 We predict the dilution using simulated data samples generated
 with the {\sc pythia} Monte Carlo~\cite{PYTHIA}.
% and a full simulation of the CDF detector.
 Control samples of $B^-$ and $\overline{B}^0$ are used to validate
 the predictions of the simulation. 
 The effectiveness of this flavor tag increases with the $p_T$ of the
 {\Bsbar}; we find $Q = 3.5\%$ ($4.0\%$) in the hadronic (semileptonic)
 decay sample.
 The fractional uncertainty on $Q$ is approximately 25\%.
 This uncertainty is dominated by the differences between data and
 simulation for kaons found close in phase space to the {\Bsbar}~\cite{DENYS}
 and for the performance of the same-side kaon tag when applied to $B^-$.

 If both a same-side tag and an opposite-side tag are present,
 we combine the information from both tags assuming they are uncorrelated.
 The addition of the same-side kaon tag increases the effective sample
 statistics by more than a factor of three.
 
%%%%%%%%%%%%%%%%%%%%%%%%%%%%%%%%%%%%%%%%%%%%%%%%%%%%%%%%%%%%%%%%%%%%%%%%%%%%%%%%
% fit and results
%%%%%%%%%%%%%%%%%%%%%%%%%%%%%%%%%%%%%%%%%%%%%%%%%%%%%%%%%%%%%%%%%%%%%%%%%%%%%%%%

 We use an unbinned maximum likelihood fit to search for {\bs} oscillations. 
 The likelihood combines mass, decay time, decay-time resolution and
 flavor tagging information for each candidate, and includes terms for
 signal and each type of background.
 The fit is done in three stages. 
 First, a combined mass and decay-time fit is performed to separate
 signal from background and to fix mass and decay-time models.
 Combined fits for $\bs$ mass and decay width in hadronic samples
 and for decay width in the semileptonic samples
 yield measurements consistent with established values
 \cite{PDG2006}.
 Next, flavor asymmetries are measured for background components.
 The last step is a fit for $\BsBsbar$ oscillations; the mass and decay-time
 models and background asymmetries are fixed from the previous
 two stages.

% Dilutions and decay time resolutions per candidate are used in the
% likelihood fit.
 The signal PDF has the general form:
% $$
% \frac{ \partial P_{\pm}}{\partial t} (t_i) = \frac{ \Gamma_s }{2} e^{-\Gamma_s t'} \cdot 
% \left[ 1 \pm {\cal A} {\cal D}_i \cos \Delta m_s t' \right ] \otimes
% {\cal G}( t_i - t', \sigma_{t_i} ) \cdot \varepsilon(t_i)
% \label{eqn:likelihood}
% $$
 \begin{eqnarray*}
 &&{\cal S}_{\pm}(t_i,\sigma_{t_i}, {\cal D}_i) = \\
 &&\varepsilon(t_i)
 \int \frac{ \Gamma_s }{2} e^{-\Gamma_s t'}
 \left[ 1 \pm {\cal A} {\cal D}_i \cos (\Delta m_s t') \right ]
 {\cal G}( t_i - t', \sigma_{t_i} )\ dt',
 \end{eqnarray*}
 where ${\cal D}_i$ is the $i$th candidate
 dilution, and $t_i$, $\sigma_{t_i}$, $\cal G$, and $\varepsilon(t)$ have been
 defined previously. Following the method described in \cite{MOSER}, we
 fit for the oscillation amplitude $\cal A$ while fixing $\Delta m_s$ to a
 probe value. When all detector effects (${\cal D}_i, \sigma_{t_i}$) are
 calibrated, the oscillation amplitude is expected to be consistent with
 ${\cal A} = 1$ when the probe value is the true oscillation frequency,
 and consistent with ${\cal A} = 0$ when the probe value is far from
 the true oscillation frequency.
 Figure~\ref{fig:amplitudeScan} (upper) shows the fitted value of the
 amplitude as a function of the oscillation frequency. 
 The sensitivity of the measurement is defined by the maximum value of {\dms } 
 where ${\cal A}=1$ is excluded at 95\% C.L.~if
 the measured value of ${\cal A}$ were zero. 
 Our sensitivity is 25.8\,{\ips } and exceeds the combined sensitivity
 of all previous experiments~\cite{PDG2006}.
 At $\dms = 17.3~\ips $, the observed amplitude
${\cal A} = 1.03\pm 0.28 ({\rm stat.})$
 is consistent with unity, indicating that the data
 are compatible with {\BsBsbar} oscillations with that frequency, while the
 amplitude is inconsistent with zero: ${\cal A}/\sigma_{\cal A}=3.7$,
 where $\sigma_{\cal A}$ is the uncertainty on ${\cal A}$.
 The negative amplitudes measured at frequencies slightly below and slightly
 above the peak frequency are expected and are due to the finite range in
 signal decay time that is imposed by the trigger and selection criteria.
 The systematic uncertainty on ${\cal A}$ is mainly due to uncertainties
 on $\sigma_{t_i}$ and ${\cal D}_i$.
 Since the effect of these uncertainties on ${\cal A}$ and $\sigma_{\cal A}$
 are correlated, the ratio ${\cal A}/\sigma_{\cal A}$ has negligible systematic
 uncertainty.

 The significance of the potential signal is evaluated from
 $\Lrat \equiv \log [ {\cal L}^{{\cal A}=0} / {\cal L}^{{\cal A}=1}(\dms)]$,
 which is the logarithm of the ratio of likelihoods for the hypothesis
 of oscillations (${\cal A}=1$) at the probe value and the hypothesis
 that ${\cal A}=0$, which is equivalent to random production flavor tags.
%that ${\cal A}=0$, which is equivalent to $\dms=\infty$.
%The significance of the signal is evaluated from the logarithm of the
%ratio of likelihoods for the hypotheses of no mixing and mixing:
%$\Lrat = \log [ {\cal L}^{{\cal A}=0} / {\cal L}^{{\cal A}=1}(\dms)]$.
 Figure~\ref{fig:amplitudeScan} (lower) shows {\Lrat } as a function of
 {\dms }. A minimal value of $\Lrat = -6.75$ is observed at
 $\dms=17.3~ \ips$.
 The significance of the signal is quantified by the
 probability that randomly tagged data would produce a value 
 of {\Lrat } lower than $-6.75$ at any value of {\dms}.
 We repeat the fit 50,000 times with random tagging decisions,
 and we find this probability is 0.2\%.
% We randomize the tag decisions in our data an measure {\Lrat}.
% We repeat this process, creating 50,000 replications of this experiment,
% each time randomizing the tag decisions in a statistically independent
% way from previous trials.
% We determine this probability to be 0.2\% by repeating the fit 50,000 times. 
% We randomize the tagging decisions in these 50,000 fits in statistically
% independent ways and count the fraction of experiments that give
% $\Lrat < -6.75$ at any value of {\dms }. 
% We determine this probability to be 0.2\% by repeating the fit many times, 
% while randomizing the tagging decisions in statistically independent  
% ways and counting the fraction of experiments that give $\Lrat < -6.75$
% at any value of {\dms }. 

 Under the hypothesis that the signal is due to {\BsBsbar} oscillations, 
 we fix ${\cal A}=1$ and fit for the oscillation frequency.
 We find {\deltaMsResult}
 and the range $17.01\,\ips < \dms < 17.84\,\ips$
 ($16.96\,\ips < \dms < 17.91\,\ips$) at 90\% (95\%) C.L. 
 All systematic uncertainties affecting ${\cal A}$ are unimportant for {\dms}. 
 The only non-negligible systematic uncertainty on {\dms } is from
 the uncertainty on the absolute scale of the decay-time measurement.
 Contributions to this uncertainty include biases in the primary-vertex
 reconstruction due to the presence of the opposite-side $b$~hadron,
 uncertainties in the silicon-detector alignment, and biases in track fitting.
 The measured {\BsBsbar} oscillation frequency is used to derive the ratio
 $\VtdVts  = \xi\sqrt{ \frac{\dmd}{\dms} \frac{m_{\Bs}}{m_{\Bd}} }$.
 As inputs we use {\massRatio}~\cite{CDF_BS_MASS} with negligible
 uncertainty, {\deltaMdPdg}\,\cite{PDG2006} and
 {\xiLat}\,\cite{XI_LAT}. We find {\VtdResult}.

%%%%%%%%%%%%%%%%%%%%%%%%%%%%%%%%%%%%%%%%%%%%%%%%%%%%%%%%%%%%%%%%%%%%%%%%%%%%%%%%
% conclusion
%%%%%%%%%%%%%%%%%%%%%%%%%%%%%%%%%%%%%%%%%%%%%%%%%%%%%%%%%%%%%%%%%%%%%%%%%%%%%%%%

 In conclusion, we present the first measurement of $\dms$.
 The precision of this measurement is better than 2\%. 
 The value of $\dms$ is consistent with standard model expectations
 \cite{CKMUT} and with previous bounds.
 Our measured value of $\dms$ allows us to determine {\VtdVts} with
 unprecedented precision and can be used to improve constraints on the
 unitarity of the CKM matrix and on scenarios involving new physics.

%%%% ACKNOWLEDGEMNTS %%%%%%%%%%%%%%%%%%%%%%%%%%%%%%%%%%%%%%%%%%%%%%%%%%%%%%%%%%%

We thank the Fermilab staff and the technical staffs of the participating
institutions for their vital contributions. This work was supported by the
U.S. Department of Energy and National Science Foundation; the Italian Istituto
Nazionale di Fisica Nucleare; the Ministry of Education, Culture, Sports,
Science and Technology of Japan; the Natural Sciences and Engineering Research
Council of Canada; the National Science Council of the Republic of China; the
Swiss National Science Foundation; the A.P. Sloan Foundation; the
Bundesministerium f\"ur Bildung und Forschung, Germany; the Korean Science and
Engineering Foundation and the Korean Research Foundation; the Particle Physics
and Astronomy Research Council and the Royal Society, UK; the Russian Foundation
for Basic Research; the Comisi\'on Interministerial de Ciencia y
Tecnolog\'{\i}a, Spain; in part by the European Community's Human Potential
Programme under contract HPRN-CT-2002-00292; and the Academy of Finland.

%%%%%%%%%%%%%%%%%%%%%%%%%%%%%%%%%%%%%%%%%%%%%%%%%%%%%%%%%%%%%%%%%%%%%%%%%%%%%%%%
% figures
%%%%%%%%%%%%%%%%%%%%%%%%%%%%%%%%%%%%%%%%%%%%%%%%%%%%%%%%%%%%%%%%%%%%%%%%%%%%%%%%

\begin{figure}[htb]
\begin{center}
\includegraphics[width=1\linewidth]{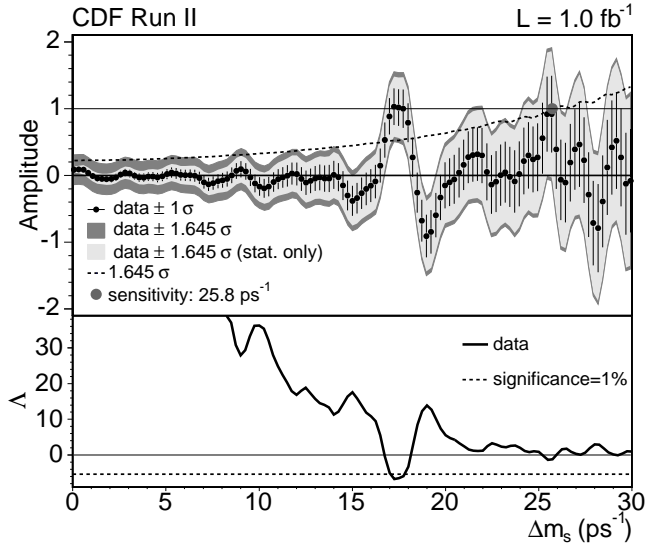}
\caption[] {
 (Upper)
 The measured amplitude values and uncertainties versus the {\BsBsbar}
 oscillation frequency $\dms$. At 17.3~$\ips$, the amplitude
 is consistent with one and inconsistent with zero at 3.7
 standard deviations. Shown in light gray and dark gray are the 95\%
 one-sided confidence level bands for statistical uncertainties only and
 including systematic uncertainties, respectively.\\
 (Lower)
 The logarithm of the ratio of likelihoods for amplitude equal to
 zero and amplitude equal to one,
 $\Lambda = \log [ {\cal L}^{{\cal A}=0} / {\cal L}^{{\cal A}=1}(\dms) ]$,
 versus the oscillation frequency.
 The deepest minimum is $\dms=17.3~\ips$,
 where $\Lambda = -6.75$. 
 The dashed horizontal line indicates the value of $\Lambda$ that corresponds
 to a probability of 1\% in the case of randomly tagged data. 
}
\label{fig:amplitudeScan}
\end{center}
\end{figure}

%%%%%%%%%%%%%%%%%%%%%%%%%%%%%%%%%%%%%%%%%%%%%%%%%%%%%%%%%%%%%%%%%%%%%%%%%%%%%%%%
% references
%%%%%%%%%%%%%%%%%%%%%%%%%%%%%%%%%%%%%%%%%%%%%%%%%%%%%%%%%%%%%%%%%%%%%%%%%%%%%%%%

%%%\input{bibliography.tex}

%%%%%%%%%%%%%%%%%%%%%%%%%%%%%%%%%%%%%%%%%%%%%%%%%%%%%%%%%%%%%%%%%%%%%%%%%%%%%%%%

\begin{thebibliography}{99}

\bibitem{MIXING}
%%
  C.~Gay, Annu.~Rev.~Nucl.~Part.~Sci.~{\bf 50}, 577 (2000).
%  We neglect a width
%  difference between $B^0_{q,H}$ and $B^0_{q,L}$.
  We set $\hbar = c = 1$ and report
  $\Delta m_q = m_{B^0_{q,H}} - m_{B^0_{q,L}}$ in inverse picoseconds.

\bibitem{CKM}
%%
  N.~Cabibbo,
  Phys.~Rev.~Lett.~{\bf 10}, 531 (1963);
  M.~Kobayashi and T.~Maskawa,
  Prog.~Theor.~Phys.~{\bf 49}, 652 (1973).

\bibitem{PDG2006}
%%
  S. Eidelman {\it et al.},
  Phys.~Lett.~B~{\bf 592}, 1 (2004) and 2005 partial update
  for the 2006 edition available on the PDG WWW pages
  (http://pdg.lbl.gov/).
  
%% \bibitem{UT}
%% %%
%%   L.-L.~Chau and W.-Y.~Keung,
%%   Phys.~Rev.~Lett.~{\bf 53}, 1802 (1984);
%%   C.~Jarlskog and R.~Stora,
%%   Phys.~Lett.~B {\bf 208}, 268 (1988);
%%   J.D.~Bjorken,
%%   Phys.~Rev.~D {\bf 39}, 1396 (1989).

\bibitem{DMD}
%%
  K.~Abe {\it et al.} (BELLE Collaboration),
  Phys.~Rev.~D {\bf 71}, 072003 (2005);
  {\bf 71}, 079903(E) (2005);
  N.~C.~Hastings {\it et al.} (BELLE Collaboration),
  Phys.~Rev.~D~{\bf 67}, 052004 (2003);
  B.~Aubert {\it et al.} (BABAR Collaboration),
  Phys.~Rev.~Lett.~{\bf 88}, 221803 (2002). 

\bibitem{DMS}
%%
  J.~Abdallah {\it et al.} (DELPHI Collaboration),
  Eur.~Phys.~J.~C {\bf 35}, 35 (2004);
  K.~Abe {\it et al.} (SLD Collaboration),
  Phys.~Rev.~D~{\bf 67}, 012006 (2003);
  A.~Heister {\it et al.} (ALEPH Collaboration),
  Eur.~Phys.~J.~C {\bf 29}, 143 (2003).


\bibitem{D0-BSMIX-2006}
%%
  V.~M.~Abazov {\it et al.} (D\O\ Collaboration),
  ``First direct two-sided bound on the $B_s^0$
  oscillation frequency,''
  hep-ex/0603029, submitted to Physical Review Letters.

\bibitem{NOTATION}
%%
  The symbol {\bs} refers to the combination of
  {\Bsbar} and {\Bs} decays.
  
\bibitem{CHARGECONJUGATE}
%%
  References to a particular process imply that
  the charge conjugate process is included as well.

%\bibitem{CDF-BMASS}
%%%
%  A.~Abulencia {\it et al.} (CDF Collaboration),
%  hep-ex/0508022.

%\bibitem{PDG2002}
%%
%  K.~Hagiwara {\it et al.},
%  Phys.~Rev.~D~{\bf 66}, 010001 (2002).

%% \bibitem{PDG2004}
%% %%
%%   S. Eidelman {\it et al.},
%%   Phys.~Lett.~B~{\bf 592} (2004) 1.

\bibitem{DETECTOR_REFERENCE}
%%
%%A more complete overview of the CDF\,II detector may be found in
%%%the J/Psi paper
  D.~Acosta {\it et al.} (CDF Collaboration),
  Phys.~Rev.~D~{\bf 71}, 032001 (2005);
%%%the CDF II TDR
  R.~Blair {\it et al.} (CDF Collaboration),
  ``The CDF-II Detector Technical Design Report,''
  FERMILAB--PUB--96--390--E (1996).

\bibitem{MUONS}
%%  
  A.~Abulencia {\it et al.} (CDF Collaboration), ``Measurements of
  inclusive W and Z cross sections in $p \overline{p}$ collisions at
  $\sqrt{s}$ = 1.96~TeV,'' hep-ex/0508029, submitted to Physical Review
  D.

\bibitem{SVX}
%%
  A.~Sill {\it et al.}, Nucl.~Instrum.~Methods Phys.~Res., Sect.~A
  {\bf 447}, 1 (2000).

\bibitem{ISL}
%%
  A.~Affolder {\it et al.}, Nucl.~Instrum.~Methods Phys.~Res., Sect.~A
  {\bf 453}, 84 (2000).

\bibitem{L00}
%%
  C.~S.~Hill {\it et al.}, Nucl.~Instrum.~Methods Phys.~Res., Sect.~A
  {\bf 530}, 1 (2004).

\bibitem{COT}
%%
  T. Affolder {\it et al.}, Nucl.~Instrum.~Methods Phys.~Res., Sect.~A
  {\bf 526}, 249 (2004).

%\bibitem{COORDINATES}
%%
%  We use a cylindrical coordinate system with $z$ pointing along the momentum
%  of the protons. Pseudorapidity is $\eta=-\ln\tan(\theta/2)$, where
%  $\theta$ is the polar angle of the momentum vector with respect to the
%  $z$~axis.

\bibitem{TOF}
%%
%  A.~Abulencia {\it et al.} (CDF Collaboration),
%  ``Direct Search for Dirac Magnetic Monopoles in $p\bar{p}$ Collisions
%  at $\sqrt{s} = 1.96$\,TeV,'' hep-ex/0509015,
%  accepted by Physical Review Letters;
%  C.~Grozis {\it et al.}, Int.~J.~Mod.~Phys.~A, {\bf 16S1C}, 1119 (2001).
   S.~Cabrera {\it et al.},
   Nucl.~Instrum.~Methods Phys.~Res., Sect.~A {\bf 494}, 416 (2002).

%\bibitem{XFT}
%%
%  E.~J.~Thomson {\it et al.},
%  IEEE Trans.~Nucl.~Sci. {\bf 49}, 1063 (2003).

\bibitem{SVT}
%%
  W.~Ashmanskas {\it et al.},
  Nucl.~Instrum.~Methods Phys.~Res., Sect.~A {\bf 518}, 532 (2004).

\bibitem{CDF-BR-BSDSPI}
%%
  A.~Abulencia {\it et al.} (CDF Collaboration),
  Phys.~Rev.~Lett.~{\bf 96}, 191801 (2006).

\bibitem{GIURGIU_THESIS}
%%
  G.~Giurgiu, Ph.~D.~thesis, Carnegie Mellon University,
  2005, FERMILAB-THESIS-2005-41.

\bibitem{BSUBC_PSI_E}
%%
  A.~Abulencia {\it et al.} (CDF Collaboration),
  ``Measurement of the $B^+_c$ Meson Lifetime using
  $B^+_c\rightarrow J/\psi e^+ \nu_e$,'' hep-ex/0603027,
  submitted to Physical Review Letters.

\bibitem{DEFINE_PT}
%%
  The transverse momentum $p_T$ is the magnitude of the component
  of the momentum perpendicular to the proton beam direction.

\bibitem{DPRODUCTION_PRL}
%%
  D.~Acosta {\it et al.} (CDF Collaboration),
  Phys.~Rev.~Lett.~{\bf 91}, 241804 (2003).

\bibitem{OWEN_PRD}
%%
  F.~Abe {\it et al.} (CDF Collaboration),
  Phys.~Rev.~D~{\bf 60}, 072003 (1999).

%\bibitem{NEUROBAYES}
%%%
%  M.~Feindt,
%  {\em A Neural Bayesian Estimator for Conditional Probability Densities},
%  e-print: physics/0402093.  

\bibitem{SST}
%%
  A.~Ali and F.~Barreiro, 
  Z.~Phys.~C {\bf 30}, 635 (1986);
  M.~Gronau, A.~Nippe, J.~L.~Rosner,
  Phys.~Rev.~D {\bf 47}, 1988 (1993);
  M.~Gronau and J.~L.~Rosner,
  Phys.~Rev.~D {\bf 49}, 254 (1994).

\bibitem{PETAR}
%%
  F.~Abe {\it et al.} (CDF Collaboration),
  Phys.~Rev.~D~{\bf 59}, 032001 (1999).

\bibitem{CDF_SIN2BETA}
%%
  T.~Affolder {\it et al.} (CDF Collaboration),
  Phys.~Rev.~D~{\bf 61}, 072005 (2000).

\bibitem{PYTHIA}
%%
  T.~Sj\"ostrand {\it et al.},
  %P.~Ed\'en, C.~Friberg, L.~L\"onnblad, G.~Miu, S.~Mrenna and E.~Norrbin,
  Computer Phys. Commun.~{\bf 135}, 238 (2001).
  We use version 6.216.

%\bibitem{BSMIXING-PRD}
%%%
%  Reference to our planned PRD on the measurement of {\Bs} oscillations.

\bibitem{DENYS}
%%
  D.~Usynin, Ph.~D.~thesis, University of Pennsylvania,
  2005, FERMILAB-THESIS-2005-68.

\bibitem{MOSER}
%%
  H.G.~Moser and A.~Roussarie,
  Nucl.~Instrum.~Methods Phys.~Res., Sect.~A {\bf A384}, 491 (1997).

\bibitem{CDF_BS_MASS}
%%
  D.~Acosta {\it et al.} (CDF Collaboration)
  Phys.~Rev.~Lett.~{\bf 96}, 202001 (2006).

\bibitem{XI_LAT}
%%
  M.~Okamoto, PoS LAT2005 (2005) 013, (hep-lat/0510113).

\bibitem{CKMUT}
%%
  M.~Bona {\it et al}. (UTfit Collaboration),
  JHEP {\bf 0507} 028 (2005);
%%    JHEP {\bf 0603} 080 (2006);
  J.~Charles {\it et al}. (CKMfitter Collaboration),
  Eur.~Phys.~J.~C {\bf 41}, 1 (2005).

\end{thebibliography}
\end{document}
%%%%%%%%%%%%%%%%%%%%%%%%%%%%%%%%%%%%%%%%%%%%%%%%%%%%%%%%%%%%%%%%%%%%%%%%%%%%%%%%